\documentclass[conference]{IEEEtran}
\usepackage{amsmath}    
\usepackage{graphicx}   
\usepackage{graphics}   
\usepackage{verbatim}   
\usepackage{color}      
\usepackage{subfigure}  
\usepackage{array}
\usepackage{amssymb}
\usepackage{epstopdf}
\newtheorem{theorem}{Theorem}[section]
\newtheorem{proposition}[theorem]{Proposition}

\newtheorem{lemma}[theorem]{Lemma}

\newtheorem{example}{Example}

\begin{document}

\title{Quantum Synchronizable Codes From Quadratic Residue Codes and Their Supercodes}

\author{\IEEEauthorblockN{Yixuan Xie$^1$, Jinhong Yuan$^1$, and Yuichiro Fujiwara$^2$}
\IEEEauthorblockA{${}^1$ School of Electrical Engineering and Telecommunications, The University of New South Wales, Sydney, Australia\\}
\IEEEauthorblockA{${}^2$ Division of Physics, Mathematics and Astronomy, California Institute of Technology, Pasadena, California 91125, USA\\
Email: Yixuan.Xie@student.unsw.edu.au, J.Yuan@unsw.edu.au, yuichiro.fujiwara@caltech.edu}
}

\maketitle

\begin{abstract}
Quantum synchronizable codes are quantum error-correcting codes designed to correct the effects of both quantum noise and block synchronization errors.
While it is known that quantum synchronizable codes can be constructed from cyclic codes that satisfy special properties,
only a few classes of cyclic codes have been proved to give promising quantum synchronizable codes.
In this paper, using quadratic residue codes and their supercodes,
we give a simple construction for quantum synchronizable codes whose synchronization capabilities attain the upper bound.
The method is applicable to cyclic codes of prime length.
\end{abstract}

\section{Introduction}

Quantum information theory has experienced remarkable progress towards developing and realizing large scale quantum computation and quantum communication.
One of the most important early landmarks in quantum information science is the discovery of quantum error-correcting codes \cite{Shor:1995,Steane1996}, which enable us to suppress decoherence and other unwanted changes to quantum states that store quantum information. Because quantum bits, or \emph{qubits}, that carry quantum information are highly vulnerable to quantum noise, various types of noise model and specialized quantum error-correcting codes have extensively been studied \cite{Lidar:2013book}. For instance, among the widely studied are Pauli errors and qubit loss \cite{Nielsen:2000book}, which roughly correspond to additive noise and erasures in classical information theory respectively.

Misalignment in block synchronization is another type of error that causes catastrophic failure, where the information processing device misidentifies the boundaries of an information block. For instance, assume that each chunk of information is encoded into a block of consecutive three bits in a stream of bits $b_i$ so that the data has a frame structure. If four blocks of information are encoded, we have twelve ordered bits $(b_0, b_1, b_2,  b_3, b_4, b_5, b_6, b_7, b_8, b_9, b_{10}, b_{11})$ in which each of the four blocks $(b_0, b_1, b_2)$,  $(b_3, b_4, b_5)$, $(b_6, b_7, b_8)$, and $(b_9, b_{10}, b_{11})$ forms an information chunk. If, for example, misalignment occurs to the right by two bits when attempting to retrieve the second block of information,
the device will wrongly read out $b_5$, $b_6$, and $b_7$ instead of the correct set of bits $b_3$, $b_4$, and $b_5$.
The same kind of error in block synchronization may be considered for a stream of qubits.

A \emph{quantum synchronizable code} is a coding scheme that corrects general quantum noise represented by Pauli errors as well as block synchronization errors. A theoretical framework of quantum synchronizable coding was first introduced in \cite{Fujiwara:Phy2013} as a quantum analogue of synchronizable coding in classical coding theory that attempts to correct both bit flips and block synchronization errors \cite{Bose:1967}. Subsequent studies have improved the original construction method and given further examples of quantum synchronizable codes \cite{Fujiwara:phy2013b,Fujiwara:2014}.

While we now have a theoretical framework of quantum synchronizable coding, there are only a few infinite classes of quantum synchronizable codes in the literature. In general, quantum synchronizable codes can be constructed from classical cyclic codes with additional properties through a method similar to the one studied in \cite{Grassl:2000}. However, while the quantum analogue of cyclic codes given in \cite{Grassl:2000} only requires a cyclic code that contains its dual, the known general construction for quantum synchronizable codes requires a chain of three cyclic codes satisfying further complicated properties, making it harder to explicitly construct promising examples.

In this work, we construct quantum synchronizable codes by exploiting special classical cyclic codes over the finite field $\mathbb{F}_2$ of order $2$, called quadratic residue codes. Quadratic residue codes tend to have large minimum distances\cite{MacWilliams:book1978}. Thus, it is reasonable to expect that quantum error-correcting codes that exploit quadratic residue codes possess good error correction performance. We show that quantum synchronizable codes from quadratic residue codes also have good block synchronization capabilities. In fact, our codes attain the known upper bound on the maximum tolerable magnitude of misalignment in some cases. Note that the concept of the proposed method also applies to the general cyclic codes.

In Section II, we review basic facts in classical coding theory. Section III revisits the framework of quantum synchronizable coding and briefly explains the procedure of synchronization recovery. Section IV discusses our construction for quantum synchronizable codes based on quadratic residue codes. Concluding remarks are given in Section V.

\section{Preliminaries}
Here we briefly review basic facts in classical coding theory. Their proofs can be found in a standard textbook in coding theory such as \cite{MacWilliams:book1978}.

A \emph{binary linear} code $[n, k, d]$ is a $k$-dimensional subspace $\mathcal{L}$ of the $n$-dimensional vector space $\mathbb{F}^{n}_2$ such that $\min\{\operatorname{wt}(\mathbf{v}) \mid \mathbf{v} \in \mathcal{L}, \mathbf{v} \not= \mathbf{0}\}=d$, where $\operatorname{wt}(\mathbf{v})$ is the number of nonzero entries in $\mathbf{v}$. A binary vector in the subspace is a \emph{codeword}. Throughout this paper, we only consider binary codes and omit the term binary.

A \emph{cyclic code} $\mathcal{C}$ of \emph{length} $n$, \emph{dimension} $k$, and \emph{minimum distance} $d$ is a linear code $[n,k,d]$ such that every cyclic shift of any codeword $\left(c_0, c_1,\ldots,c_{n-1}\right) \in \mathcal{C}$ is also a codeword of $\mathcal{C}$. When the focus is on the length and dimension of a code, we may drop the third argument from the parameter notation. The \emph{dual code} $\mathcal{C}^{\perp}$ of $\mathcal{C}$ is defined as $$\mathcal{C}^{\perp} = \{\mathbf{c}'\in\mathbb{F}_2^n \mid \mathbf{c}'\cdot\mathbf{c} = 0 \mbox{\ for all\ }\mathbf{c}\in\mathcal{C}\}.$$
The dimension $\operatorname{dim}\left(\mathcal{C}^{\perp}\right)$ of $\mathcal{C}^{\perp}$ is $n-\operatorname{dim}\left(\mathcal{C}\right)=n-k$.
Trivially, the dual code of a cyclic code is also a cyclic code.

By regarding a codeword as a coefficient vector of a polynomial in $\mathbb{F}_{2}[x]$, an $[n,k]$ cyclic code $\mathcal{C}$ generated by the unique monic polynomial $g(x)$ of minimum degree can be seen as the principal ideal $\langle g(x)\rangle$ closed by the ring $\mathbb{F}_{2}[x]/\left(x^n - 1\right)$.
This unique polynomial $g(x)$ is called the \emph{generator polynomial} of the cyclic code. Its degree ${\operatorname{deg}\left(g(x)\right)}$ is $n-k$. We may understand the cyclic code as
$$\mathcal{C}=\{m(x)g(x) \mid m(x) \in \mathbb{F}_{2}[x], \operatorname{deg}(m(x))< k\}.$$
The polynomial $h(x)$ such that $g(x)h(x)= x^n-1$ is called the \emph{check polynomial} of $\mathcal{C}$.
The dual code $\mathcal{C}^{\perp}$ has generator polynomial $g^{\perp}(x)$ of the form
\begin{align}
\label{equ:genDual}
g^{\perp}(x) = x^{\operatorname{deg}(h(x))}h(x^{-1}).
\end{align}

Let $\mathcal{C}_1=\langle g_1 (x)\rangle$ and $\mathcal{C}_2=\langle g_{2} (x)\rangle$ be two cyclic codes of length $n$. If $\mathcal{C}_2\subseteq\mathcal{C}_1$, that is, if $\mathcal{C}_1$ contains all codewords of $\mathcal{C}_2$, then the generator polynomial $g_1(x)$ divides every codeword of $\mathcal{C}_2$, which means that for every $c(x)\in \mathcal{C}_2$ there exits a polynomial $f_c(x)$ of degree $\operatorname{deg}(c(x))-\left(n-\operatorname{dim}\left(\mathcal{C}_1\right)\right)$ such that $c(x)=f_c(x)g_1(x)$ in $\mathbb{F}_{2}[x]$.
The smaller code $\mathcal{C}_2$ is a \emph{subcode} of $\mathcal{C}_1$ while $\mathcal{C}_1$ is a \emph{supercode} of $\mathcal{C}_2$.
A cyclic code is \emph{dual-containing} if it is a supercode of its dual code.

\section{Quantum Synchronizable Code}

In this section, we review the framework of quantum synchronizable codes and the synchronization recovery procedures. The basic notions and proofs of the facts in quantum information theory we use in this section can be found in \cite{Nielsen:2000book}. For complete treatments of block synchronization in the context of quantum information, the interested reader is referred to the original articles \cite{Fujiwara:Phy2013,Fujiwara:phy2013b}.

\subsection{Quantum Synchronizable Codes}
An $[[n, k]]$ \emph{quantum error-correcting code} is a coding scheme that encodes $k$ logical qubits into $n$ physical qubits. As in the classical case, $n$ and $k$ are the \emph{length} and \emph{dimension} of the code, respectively. Typically, quantum error-correcting codes are designed to correct the effects of bit errors and phase errors caused by Pauli operators $X$ and $Z$ respectively under the assumption that both bit error due to $X$ and phase error due to $Z$ may occur on the same qubit. A $(c_l, c_r)$-$[[n, k]]$ \emph{quantum synchronizable code} is an $[[n, k]]$ quantum error-correcting code that corrects not only bit errors and phase errors but also misalignment to the left by $c_l$ qubits and to the right by $c_r$ qubits for some nonnegative integers $c_l$ and $c_r$.

The general construction method for quantum synchronizable codes developed in \cite{Fujiwara:Phy2013,Fujiwara:phy2013b} employs a notion in finite algebra. Let $f(x) \in \mathbb{F}_{2}[x]$ be a polynomial over $\mathbb{F}_{2}$ such that $f(0)=1$.
The \emph{order} $\operatorname{ord}\left(f(x)\right)$ of the polynomial $f(x)$ is the cardinality $\left\vert\{x^a \pmod{f(x)}\ \middle\vert\ a \in \mathbb{N}\}\right\vert$, where $\mathbb{N}$ is the set of positive integers.

\begin{theorem}
\label{them1}\cite{Fujiwara:phy2013b}
Let $\mathcal{C}_1=\langle g_1(x)\rangle$ and $\mathcal{C}_2=\langle g_2(x)\rangle$ be two cyclic codes of parameters $[n,k_1,d_1]$ and $[n,k_2,d_2]$ with $k_1>k_2$
respectively such that $\mathcal{C}_2 \subset \mathcal{C}_1$ and $\mathcal{C}_2^{\perp}\subseteq\mathcal{C}_2$.
Define $f(x)$ of degree $k_1-k_2$ to be the quotient of $\frac{g_2(x)}{g_1(x)}$ over $\mathbb{F}_2[x]/(x^n-1)$.
For any pair of nonnegative integers $c_l, c_r$ satisfying $c_l+c_r<\operatorname{ord}\left(f(x)\right)$, there exists a $(c_l, c_r)$-$[[n+c_l + c_r, 2k_2 -n]]$ quantum synchronizable code that corrects at least up to $\lfloor\frac{d_1-1}{2}\rfloor$ bit errors and at least up to $\lfloor\frac{d_2-1}{2}\rfloor$ phase errors.
\end{theorem}

Theorem \ref{them1} requires a pair of cyclic codes $\mathcal{C}_1, \mathcal{C}_2$ of the same length and dimension $k_1 > k_2 > \lceil\frac{n}{2}\rceil$ to construct a quantum synchronizable code of positive dimension. To design a good quantum synchronizable code, it is generally desirable to choose cyclic codes with good minimum distances while ensuring $\operatorname{ord}\left(f(x)\right)$ to be as large as possible. In addition to these criteria, the cyclic codes must satisfy the chain condition that $\mathcal{C}_2^{\perp}\subseteq\mathcal{C}_2\subset\mathcal{C}_1$. Note that this is stronger than the dual-containing condition for the quantum cyclic codes given in \cite{Grassl:2000}. In what follows, when a pair of cyclic codes are written as $\mathcal{C}_1$ and $\mathcal{C}_2$, we always assume that they satisfy the chain condition and that their generator polynomials are $g_1(x)$ and $g_2(x)=f(x)g_1(x)$ for some polynomial $f(x)$, respectively.

\subsection{Encoding}
Since $\operatorname{dim}\left(\mathcal{C}_2\right)=k_2$ and $\operatorname{dim}\left(\mathcal{C}_2^{\perp}\right)=n-k_2$, the dimension of cosets is $\operatorname{dim}\left(\mathcal{C}_2/\mathcal{C}^{\perp}_2\right)=k_2-n+k_2=2k_2-n$. Hence, the number of cosets is $2^{2k_2-n}$. Let $\mathcal{B}=\{b_i(x)\mid0<i\leq 2^{2k_2-n}\}$ be a system of representatives of the cosets. Then the set
$$
\label{equ:orthonormalstates}
V = \left\{\left|\mathcal{C}_2^{\perp} + b_i(x)\right\rangle \ \middle\vert\ b_i(x)\in\mathcal{B}\right\}
$$
of $2^{2k_2-n}$ quantum states forms an orthogonal basis of a vector space of dimension $2^{2k_2-n}$,
where
$$\left|\mathcal{C}_2^{\perp} + b_i(x)\right\rangle= \frac{1}{\sqrt{\vert\mathcal{C}_2^{\perp}\vert}}\sum_{c(x) \in \mathcal{C}_2^{\perp}}\left|c(x) + b_i(x)\right\rangle.$$

Take an arbitrary $2^{2k_2-n}$-qubit state $\vert\psi\rangle$ to be encoded.
Using the standard encoder for Calderbank-Shor-Steane (CSS) codes \cite{Nielsen:2000book}, the state $|\psi\rangle$ is transformed into $n$-qubit state $\vert\psi\rangle_{enc}=\sum_i \alpha_i \vert v_i \rangle$, where $v_i \in V$.

Recall that $g_1(x)$ is the generator polynomial of $\mathcal{C}_1$. Apply the unitary operator $U_g$ that adds the coefficient vector of $g_1(x)$:
$$U_g\vert\psi\rangle_{enc}\rightarrow\sum_i \alpha_i \vert v_i + g_1\rangle.$$

Let $c_r$, $c_l$ be nonnegative integers such that $c_l+c_r < \operatorname{ord}\left(f(x)\right)$.
By attaching extra $c_l$ and $c_r$ ancilla qubits to the left and to the right of the original state respectively
and then applying CNOT gates, the state is taken to the final encoded $(n+c_l+c_r)$-qubit state
$$\vert 0\rangle^{\otimes c_l}U_g(\vert\psi\rangle_{enc})\vert0\rangle^{\otimes c_r}\rightarrow\sum_i \alpha_i\vert l_i, v_i+g_1, r_i \rangle=\vert\Psi\rangle_{enc},$$
where $l_i$ and $r_i$ are the last $c_l$ and the first $c_r$ portions of the vector $v_i+g_1$, respectively.

\subsection{Synchronization Recovery}
Here we give a brief overview of the procedures for error correction and synchronization recovery. For the mathematical details, see \cite{Fujiwara:Phy2013,Fujiwara:phy2013b}.

Assume that the device gathered qubits of one block length, that is, consecutive $n+c_l + c_r -1$ qubits, and tries to correct errors caused by Pauli operators and misalignment if necessary. Let $\mathcal{T} = (t_{0}, t_{1}, \ldots, t_{n+c_l + c_r -1})$ be the collection of $n+c_l+c_r$ qubits at the output of the quantum channel. If block synchronization is correct, $\mathcal{T}$ forms a properly aligned block encoded as $\vert\Psi\rangle_{enc}$. We assume that $\mathcal{T}$ may be misaligned by $\theta$ qubits to the right, where $-c_l \leq \theta \leq c_r$. When $\theta$ is negative, it means that misalignment is to the left by $\vert \theta \vert$ qubits.

Let $\mathcal{S} = (s_{0}, s_{1}, \ldots, s_{n+c_l + c_r -1})$ be the $n+c_l+c_r$ qubits of $\vert\Psi\rangle_{enc}$. The device first focuses on consecutive $n$ qubits $\mathcal{W} = (t_{c_l}, t_{c_l+1}, \ldots, t_{c_l+n-1})$ in the middle of $\mathcal{T}$. Because of the potential misalignment, this set of qubits is $\mathcal{W} = (s_{c_l+\theta}, s_{c_l+1+\theta}, \ldots, s_{c_l+n-1+\theta})$.

Let $E$ be the $(n+c_l+c_r)$-fold tensor product of single Pauli operators that represents errors that occurred on $\vert\Psi\rangle_{enc}$. The corrupted state at the quantum channel output is given by
\begin{align}
E\vert\Psi\rangle_{enc} = \sum_i \alpha_i(-1)^{(l_i, v_i+g_1, r_i)\cdot\mathbf{e_p}} \vert (l_i, v_i+g_1, r_i) + \mathbf{e_b}\rangle, \notag
\end{align}
where $\mathbf{e_b}$ and $\mathbf{e_p}$ are binary vectors representing bit and phase errors, respectively.

The device first corrects bit errors on $\mathcal{W}$ and then detects misalignment. Let $\mathcal{H}_{\mathcal{C}_1}$ be the full-rank parity-check matrix of $\mathcal{C}_1$ used for encoding. Using the stabilizer generators defined by $\mathcal{C}_1$, the decoding circuit obtains the syndrome for bit errors as in the standard two-step decoding of a CSS code:
\begin{equation*}
\label{equ:biterrocorrection}
E|\Psi\rangle_{enc}|0\rangle^{\otimes n-k_1}\rightarrow E|\Psi\rangle_{enc}|\mathbf{e_b}\mathcal{H}_{\mathcal{C}_1}^T\rangle.
\end{equation*}
If the number of bit errors is at most $\lfloor\frac{d_1-1}{2}\rfloor$ in $\mathcal{W}$, applying $X$ Pauli operators to the qubits specified by $\mathbf{e_b}\mathcal{H}_{\mathcal{C}_1}^T$ eliminates all bit errors within the window $\mathcal{W}$.

The next step is to identify how many qubits away $\mathcal{W}$ is from the correct position $\mathcal{S}$, that is, identifying the magnitude $\theta$. To this end, we manipulate the polynomials used as the labels of each basis state. Such operations can be done, for example, by a quantum shift register given in \cite{Grassl:2000}.

Note that the condition that $\mathcal{C}_2^{\perp}\subset\mathcal{C}_2\subset\mathcal{C}_1$ implies that any codeword $c^{\perp}_i(x)\in\mathcal{C}_2^{\perp}$ also belongs to $\mathcal{C}_2$ and $\mathcal{C}_1$. Hence, each basis of the state of $\mathcal{S}$ is a sum of states of the form
\begin{equation*}
\label{equ:synchrorecovery}
\begin{array}{l}
\left\vert c_i^ \bot (x) + {b_i}(x) + {g_1}(x)\right\rangle =\\
\begin{array}{*{20}{c}}
{}&{}&{}&{\left\vert{v_1}(x)f(x){g_1}(x) + {v_2}(x)f(x){g_1}(x) + {g_1}(x)\right\rangle,}
\end{array}
\end{array}
\end{equation*}
for some polynomials $v_1(x)$ and $v_2(x)$ whose degrees are less than $k_2$. Because of the misalignment, each basis of the state of $\mathcal{W}$ is a linear combination of states of the form $\left\vert x^\theta\left(c_i^ \bot (x) + {b_i}(x) + {g_1}(x)\right)\right\rangle$. Thus, the quotient of the label of each basis of the state of $\mathcal{W}$ divided by $g_1(x)$ is $x^\theta({v_1}(x)f(x) + {v_2}(x)f(x)+1)$. Dividing this quotient by $f(x)$ gives $x^\theta$ as the remainder. Thus, if $c_l+c_r < \operatorname{ord}(f(x))$, the synchronization error $\theta$ is uniquely determined.

Because we identified the magnitude and direction of misalignment, synchronization can be recovered. Possible bit errors on qubits outside $\mathcal{W}$ and phase errors can be corrected by treating the quantum synchronizable code as a CSS code. For more details on the final steps, we refer the reader to \cite{Fujiwara:Phy2013}.

\section{Use of Cyclic Supercodes}

We now provide a construction for quantum synchronizable codes designed from special cyclic codes. Recall that Theorem \ref{them1} requires a pair of cyclic codes $\mathcal{C}_1$, $\mathcal{C}_2$ of parameters $[n,k_1]$ and $[n,k_2]$ with $k_1>k_2$ that satisfy the condition $\mathcal{C}_2^{\perp}\subseteq\mathcal{C}_2\subset\mathcal{C}_1$. We first construct a cyclic code $\mathcal{C}_2$ in such a way that its dual code $\mathcal{C}_2^{\perp}$ is a subspace of $\mathcal{C}_2$. We then obtain a cyclic supercode $\mathcal{C}_1$ by inserting codewords into $\mathcal{C}_2$. While we only apply this idea to a small, specific class of cyclic codes, this general principle of producing supercodes may be applicable to other cyclic codes to ensure good synchronization recoverability if code lengths are primes.

\subsection{Dual-containing Cyclic Codes: $\mathcal{C}_2^{\perp}\subset\mathcal{C}_2$}

Let $p$ be a prime of the form $p \equiv \pm 1 \pmod{8}$. Define $\mathcal{Q^R}=\{x^2 \pmod{p} \mid 1\leq x\leq \frac{p-1}{2}\}$ and $\mathcal{Q^{NR}} = \{1,2,\ldots,p-1\}\backslash\mathcal{Q^R}$ to be the sets of $\frac{p-1}{2}$ nonzero quadratic residues and $\frac{p-1}{2}$ quadratic non-residues respectively.
Take a primitive $p$-th root $\alpha$ of unity in $\mathbb{F}_{2^t}$, where $t$ is the smallest positive integer such that $p$ divides $2^t-1$.
Let
\begin{align*}
g_\mathcal{R}(x)=\prod_{i\in\mathcal{Q^R}} \left(x-\alpha^i \right)
 \hspace{0.2cm}\text{and} \hspace{0.2cm}
g_\mathcal{NR}(x)=\prod_{i\in\mathcal{Q^{NR}}} \left(x-\alpha^{i} \right).
\end{align*}
Note that $g_\mathcal{R}(x)$ and $g_\mathcal{NR}(x)$ are both in $\mathbb{F}_{2}[x]$.
The pair $\mathcal{C}_\mathcal{R} = \langle g_\mathcal{R}(x)\rangle$ and $\mathcal{C}_\mathcal{NR} = \langle g_\mathcal{NR}(x)\rangle $ are $[p, \frac{p+1}{2}]$ cyclic codes known as \emph{quadratic residue codes} over $\mathbb{F}_2$.
By the same token, the two polynomials
$$\bar{g}_\mathcal{R}(x)=(x-1)\prod_{i\in\mathcal{Q^R}} \left(x-\alpha^i \right)$$
and
$$\bar{g}_\mathcal{NR}(x)=(x-1)\prod_{i\in\mathcal{Q^{NR}}} \left(x-\alpha^{i} \right)$$
generate $[p, \frac{p-1}{2}]$ cyclic codes $\mathcal{\bar{C}}_\mathcal{R}=\langle \bar{g}_\mathcal{R}(x)\rangle$ and $\mathcal{\bar{C}}_\mathcal{NR}=\langle\bar{g}_\mathcal{NR}(x)\rangle$. The latter pair may also be referred to as quadratic residue codes in the literature.
It is known that quadratic residue codes tend to have large minimum distances. The following is a well-known general lower bound, known as the square root bound.
\begin{theorem}[Square Root Bound]
The minimum distance $d$ of a quadratic residue code of length $p$ is at least $\sqrt{p}$. If $p \equiv -1 \pmod{4}$, then $d^2-d+1 \geq p$.
\end{theorem}

For small quadratic residue codes, tables of exact parameters can be found in \cite{Huffman:book2003}.

To take advantage of quadratic residue codes for constructing quantum synchronizable codes, we use the fact that the larger one of each pair is dual-containing if the length is $-1$ modulo $8$. A detailed account on the properties of quadratic residue codes and their duals can be found in \cite[Ch.\ 16]{MacWilliams:book1978}.
For convenience, we give a short proof of the simple fact.
\begin{lemma}
\label{lem:QRcontainQRC}
The quadratic residue codes $\mathcal{C}_\mathcal{R}$, $\mathcal{C}_\mathcal{NR}$, $\mathcal{\bar{C}}_\mathcal{R}$ and $\mathcal{\bar{C}}_\mathcal{NR}$ of length $p \equiv -1 \pmod{8}$ have the following properties:
\begin{align*}
&1)\hspace{1cm} \mathcal{C}_\mathcal{R}^{\perp}=\mathcal{\bar{C}}_\mathcal{R}, \hspace{0.5cm} \mathcal{C}_\mathcal{NR}^{\perp}=\mathcal{\bar{C}}_\mathcal{NR}.\\
&2) \hspace{1cm} \mathcal{C}_\mathcal{R}^{\perp}\subset\mathcal{C}_\mathcal{R}, \hspace{0.5cm} \mathcal{C}_\mathcal{NR}^{\perp}\subset\mathcal{C}_\mathcal{NR}.
\end{align*}
\end{lemma}

\begin{IEEEproof}
Since $\mathcal{Q^R}$ and $\mathcal{Q^{NR}}$ are disjoint and do not contain $0$, we have
$x^p-1 = \left(x-1\right)g_\mathcal{R}(x)g_\mathcal{NR}(x)$.
The zeros of $g_\mathcal{R}(x)$ and $g_\mathcal{NR}(x)$ are $\{\alpha^i \mid i\in\mathcal{Q^R}\}$ and $\{\alpha^{i} \mid i\in\mathcal{Q^{NR}}\}$, respectively. Hence by Equation (\ref{equ:genDual}), the zeros of $\mathcal{C}_\mathcal{R}^{\perp}$ are $1$ and $\alpha^{-i}$ for $i \in \mathcal{Q^{NR}}$, and the zeros of $\mathcal{C}_\mathcal{NR}^{\perp}$ are $1$ and $\alpha^{-i}$ for $i \in \mathcal{Q^{R}}$.
Note that $\alpha^i\in\mathcal{Q^R}$ if and only if $i$ is even and that $\alpha^i\in\mathcal{Q^{NR}}$ if and only if $i$ is odd. When $p \equiv -1 \pmod{8}$, we have $i \in \mathcal{Q^R}$ if and only if $-i \in \mathcal{Q^{NR}}$. Hence, $\mathcal{C}_\mathcal{R}^{\perp}=\mathcal{\bar{C}}_\mathcal{R}$ and $\mathcal{C}_\mathcal{NR}^{\perp}=\mathcal{\bar{C}}_\mathcal{NR}$. 
Since $\mathcal{C}_\mathcal{R} = \langle g_\mathcal{R}(x)\rangle$ and $\mathcal{C}_\mathcal{R}^{\perp}=\mathcal{\bar{C}}_\mathcal{R}=\langle \bar{g}_\mathcal{R}(x)\rangle$, it is trivial that $\mathcal{C}_\mathcal{R}$ is dual-containing. By the same token, $\mathcal{C}_\mathcal{NR}$ is a dual-containing code.
\end{IEEEproof}

\begin{example}
\label{exp:QRcontainQRC}
Consider the set of nonzero quadratic residues modulo $31$
\begin{align*}
\mathcal{Q^R} &= \{1,2^2,3^2,4^2,\ldots, 15^2\} \notag\\
&=\{1, 4, 9, 16, 25, 5, 18, 2, 19, 7, 28, 20, 14, 10, 8\}.
\end{align*}
The generator polynomial of the quadratic residue code $\mathcal{C}_\mathcal{R}$ of length $p=31$ is then
\begin{align*}
\label{exp:DualQR}
g_\mathcal{R}(x) = x^{15} + x^{12}+x^7+x^6 + x^2+x+1. \notag
\end{align*}
Multiplying by $x+1$ gives the generator polynomial of $\mathcal{\bar{C}}_\mathcal{R}$
\begin{align}
\bar{g}_\mathcal{R}(x) = x^{16} +x^{15}+x^{13}+x^{12}+x^8+x^6+x^3+x+1. \notag
\end{align}
Plugging $g_\mathcal{R}(x)$ into Equation (\ref{equ:genDual}) also gives $\bar{g}_\mathcal{R}(x)$, which means that this is the generator polynomial of the dual code $\mathcal{C}_\mathcal{R}^{\perp}$ as well.
\hfill $\blacksquare$
\end{example}


\subsection{Cyclic supercodes of $\mathcal{C}_2$}

Lemma \ref{lem:QRcontainQRC} provides a $[p, \frac{p+1}{2}, d]$ dual-containing cyclic code $\mathcal{C}_2$ for prime $p \equiv-1 \pmod{8}$ with $d \geq \sqrt{p}$. To obtain another cyclic code $\mathcal{C}_1$ such that $\mathcal{C}_2\subset\mathcal{C}_1$, we increase the number of codewords by deleting a factor from the generator polynomial of $\mathcal{C}_2$ we already have. Note that, by definition, a cyclic code is a subcode of another if its generator polynomial is divisible by the other. Thus, if the generator polynomial of $\mathcal{C}_2$ has more than one factor, deletion always gives a supercode. As the following proposition shows, a particularly interesting case is when $p$ is a Mersenne prime.
\begin{proposition}
\label{proposition:Reducibleg2}
Let $\mathcal{C}=\mathcal{C}_\mathcal{R}$ be the quadratic residue code of length $p$ generated by $g_2(x)=\prod_{i\in\mathcal{Q^R}}\left(x-\alpha^i\right)$. If $p=2^l-1$,
then $g_2(x)$ can be factored into $\frac{2^{l-1}-1}{l}$ irreducible polynomials of degree $l$, that is,
\begin{equation}
\label{equ:generatorPolyInCycSets}
g_2(x)=\prod_{j} M_j(x),
\end{equation}
where $M_s(x)$ is the minimal polynomial of $\alpha^s$ over $\mathbb{F}_2$ and $\operatorname{deg}\left(M_j(x)\right)=l$ for all $j$.
\end{proposition}

The above proposition can be proved through the concept of cyclotomy \cite{MacWilliams:book1978}. For nonnegative integers $s$ and $n$, the \textit{cyclotomic coset} $C_{s,n}$ of $s$ modulo $n$ over $\mathbb{F}_2$ is the set
\[C_{s,n} = \{s2^i \mod{n} \ \vert \ i \in \mathbb{N}\}.\]
Since $C_{s,n} = C_{s',n}$ for $s' \in C_{s,n}$, we may take
\[S_n = \{\min\{t \ \vert \ t \in C_{s,n}\} \ \vert \ s \in \mathbb{N}\cup\{0\}\}\]
as a system of representatives of the cyclotomic cosets by picking the smallest element from each set. The integers modulo $n$ are partitioned into cyclotomic cosets in a way described as
\[\{0,1,\dots,n-1\} = \bigcup_{s \in S_n} C_{s,n}.\]
Let $\alpha$ be a primitive $n$th root of unity in $\mathbb{F}_{2^{\vert C_{1,n}\vert}}$. By definition, the minimal polynomial $M_s(x)$ of $\alpha^s$ over $\mathbb{F}_2$ can be expressed as
\[M_s(x) = \prod_{i \in C_{s,n}}(x-\alpha^i).\]
Note that for any $s$, the cardinality $\vert C_{s,n}\vert$ is a divisor of $\vert C_{1,n}\vert$. When $n$ is a Mersenne prime $2^l-1$,  we have $\vert C_{1,n}\vert = l$. Because $l$ is also a prime, when $n$ is a prime of the form $n=p=2^l-1$, each $C_{s,n}$ is of size $l$ as well, proving Proposition \ref{proposition:Reducibleg2}.

Because of the one-to-one correspondence between cyclic codes and monic divisors of $x^n-1$, deleting one or more factors $M_j(x)$ gives another generator polynomial that results in a cyclic code of higher dimension containing the dual-containing cyclic code $\mathcal{C}_\mathcal{R}$. Trivially, if we delete $z$ factors, the dimension of the supercode is higher than that of $\mathcal{C}_\mathcal{R}$ by $zl$. Applying Theorem \ref{them1} to this supercode as $\mathcal{C}_1$ together with the dual-containing quadratic residue code gives a quantum synchronizable code.

It is also notable that any supercode $\mathcal{C}_1$ obtained by deleting a minimal polynomial of the quadratic residue code is also dual-containing. Thus, we have a chain of cyclic codes, each of which is dual-containing itself and contains all smaller ones. Therefore, we can construct a quantum synchronizable code from any pair of codes in the chain.

\subsection{Maximum misalignment tolerance}

In the context of quantum synchronizable codes, we would like to maximize $\operatorname{ord}\left(f(x)\right)$, where $f(x)$ is the quotient in Theorem \ref{them1}, in order to tolerate as large magnitude of misalignment as possible. It is known that the maximum tolerable magnitude of a quantum synchronizable code is upper bounded by its length \cite{Fujiwara:phy2013b}. We prove that the quantum synchronizable codes from quadratic residue codes given in the previous subsection attain this bound.
\begin{lemma}
\label{corolast}
Let $\mathcal{C}_1=\langle g_1(x)\rangle$ and $\mathcal{C}_2=\langle g_2(x)\rangle$ be cyclic codes of length $p$ such that $\mathcal{C}_2 \subset \mathcal{C}_1$
and $\mathcal{C}_1 \not= \mathcal{C}_2$.
Define $f(x)$ to be the polynomial such that $g_2(x)=f(x)g_1(x)$. If $p$ is a prime, then $\operatorname{ord}\left(f(x)\right)=p$.
\end{lemma}
\begin{IEEEproof}
Because the generator polynomial of a cyclic code of length $p$ divides $x^p-1$, its factor also divides $x^p-1$. Hence, the factor $f(x)$ of $g_2(x)$ divides $x^p-1$ as well, which
implies that $x^p \equiv 1 \pmod{f(x)}$.
Hence, because $\operatorname{ord}(f(x))=\left\vert\{x^a \pmod{f(x)}\ \middle\vert\ a \in \mathbb{N}\}\right\vert$,
the order of $f(x)$ is a divisor of $p$.
Since $p$ is a prime and $f(x) \not= 1$ by assumption, we have $\operatorname{ord}(f(x))=p$ as desired.
\end{IEEEproof}

We now give our main theorem.
\begin{theorem}
Let $p = 2^l-1$ be a Mersenne prime. For nonnegative integers $c_l$, $c_r$ and $z$  such that $c_l+c_r < p$ and $z \leq \frac{2^{l-1}-l-1}{l}$, there exits a quantum synchronizable code of parameters $(c_l, c_r)$-$[[p+c_l+c_r, 2zl+1]]$.
\end{theorem}
\begin{IEEEproof}
Take a quadratic residue code of length $p= 2^l-1$ generated by the nonzero quadratic residues. By Proposition \ref{proposition:Reducibleg2}, its generator polynomial has $\frac{2^{l-1}-1}{l}$ minimal polynomials of degree $l$ as its factors. Thus we have a chain of $\frac{2^{l-1}-1}{l}$ cyclic codes in which a code contains all other smaller ones. Note that a supercode of a dual-containing code is also dual-containing. Thus, by applying Theorem \ref{them1} and Lemma \ref{corolast} the cyclic code generated by the polynomial that is obtained by deleting $z$ factors and another one obtained by deleting $z+y$ factors for some positive integer $y$, we obtain a quantum synchronizable code of desired parameters.
\end{IEEEproof}

\begin{example}
As in Example \ref{exp:QRcontainQRC}, let $p=2^{5}-1$ and take the set $\mathcal{Q^R}=\{x^2 \pmod{p} \mid 1\leq x\leq {2^{l-1}-1}\}$ of nonzero quadratic residues modulo $31$. Then $\mathcal{Q^R}$ is the union of $\frac{2^4 -1}{5}=3$ cyclotomic cosets of field $\mathbb{F}_{2^5}$ as follows.
\begin{align}
\mathcal{Q^R}&=C_{1, 31}\cup C_{5, 31}\cup C_{7, 31}, \notag
\end{align}
where $C_{1, 31} = \{1, 2, 4, 8, 16\}$, $C_{5, 31} = \{5, 10, 20, 9, 18\}$ and $C_{7, 31} = \{7, 14, 28, 25, 19\}$.

Let $\mathcal{C}_2=\langle g_\mathcal{R}(x)\rangle$. Since $g_\mathcal{R}(x)$ is the product of the minimal polynomials $M_s(x)$ of $\alpha^s$ over $\mathbb{F}_2$ for $s \in \mathcal{Q^R}$, we have
\begin{align}
g_\mathcal{R}(x) = M_1(x)M_5(x)M_7(x), \notag
\end{align}
with 
\begin{align}
&M_1(x)=x^5+x^2 +1, \notag \\
&M_5(x)=x^5+x^4+x^2+x+1, \mbox{\ and} \notag \\
&M_7(x)=x^5+x^3+x^2+x+1. \notag
\end{align}
Note that each one of $M_1(x)$, $M_5(x)$ and $M_7(x)$ divides $x^{31} - 1$. Let $\mathcal{C}_1=\langle g_1(x)=\frac{g_\mathcal{R}(x)}{f(x)}\rangle$. If we delete $z=1$ minimal polynomial, $f(x)=M_j(x)$ for $j\in\{1,5,7\}$, the dimension of $\mathcal{C}_1$ is equal to $\operatorname{dim}(\mathcal{C}_1) = p-\operatorname{deg}(g_\mathcal{R}(x))+zl=31-15+5 = 21$. If we delete $z=2$ factors, then $f(x)=M_{j_1}(x)M_{j_2}(x)$ for $j_1,j_2\in\{1, 5, 7\}$ and $j_1\neq j_2$, so the dimension of $\mathcal{C}_1$ in this case is $\operatorname{dim}(\mathcal{C}_1)=31-15+10 = 26$. In both cases, the $\operatorname{ord}\left(f(x)\right)=2^{l}-1 = p =31$.
Since $\operatorname{deg}(g_\mathcal{R}(x))=15$ and $\operatorname{dim}(\mathcal{C}_2)=16$, for arbitrary pair of non-negative integer $c_l$ and $c_r$ such that $c_l+c_r<31$, we have a $(c_l, c_r)$-$[[31+c_l+c_r, 1]]$ quantum synchronizable code.

Further, let $z=1$ and $\mathcal{C}_2=\langle g_{\mathcal{R}}(x)\rangle\subset\mathcal{C}_3$ with $\mathcal{C}_3=\langle M_{j_1}(x)M_{j_2}(x)\rangle$ for $j_1,j_2\in\{1, 5, 7\}$ and $j_1\neq j_2$. If $y=1$, by removing $z+y=2$ minimal polynomials $M_i(x)$ from $g_{\mathcal{R}}(x)$, we obtain another cyclic code $\mathcal{C}_4$ such that $\mathcal{C}_2\subset\mathcal{C}_3\subset\mathcal{C}_4$. Since $\dim(\mathcal{C}_3)=21>\lceil\frac{n}{2}\rceil$,  by Theorm \ref{them1} for $c_l+c_r<31$, $\mathcal{C}_3$ and $\mathcal{C}_4$ form a $(c_l, c_r)$-$[[31+c_l+c_r, 2zl+1=11]]$ quantum synchronizable code.
\hfill $\blacksquare$
\end{example}

\section{Conclusion}

We studied a method for constructing quantum error-correcting codes that can also recover block synchronization. Quantum synchronizable codes were derived from sets of quadratic residues by designing chains of dual-containing cyclic codes. We showed that these quantum synchronizable codes possess the highest possible tolerance against synchronization errors. Our construction is particularly flexible when the length is a Mersenne prime because we may be able to choose a pair from a long chain of dual-containing cyclic codes. This adds variety in dimension and minimum distance to the resulting quantum synchronizable codes.

Our method allows for easily calculating the lengths, dimensions, and synchronization recovery abilities of our quantum synchronizable codes. However, the exact minimum distances seem quite difficult to compute. In general. it is a very difficult problem to find the exact minimum distance of a cyclic code even for codes of modest length. While there are some known general bounds on the minimum distance of a cyclic code and quadratic residue code such as the square root bound and BCH bound \cite{MacWilliams:book1978}, the search for sharper bounds is still one of the central problems in coding theory today. Because quadratic residue codes and their variants are among the more interesting classical codes, the minimum distances and weight distributions of quadratic residue codes and their supercodes are of interest not only for the purpose of block synchronization for qubit streams but also of importance on their own right. We hope that further progress will be made in this direction in future work.



\begin{thebibliography}{1}
\bibitem{Shor:1995} P. W. Shor, ``Scheme for reducing decoherence in quantum memory,'' \emph{Phys. Rev. A}, vol. 52, pp. 2493-2496, 1995.
\bibitem{Steane1996} A. M. Steane, ``Error Correcting Codes in Quantum Theory,'' \emph{Phys. Rev. Lett.}, vol. 77, pp. 793-797, 1996.
\bibitem{Lidar:2013book} \emph{Quantum Error Correction}, D. A. Lidar and T. A. Brun, Eds. \emph{Cambridge Uni. Press}, New York, 2013.
\bibitem{Nielsen:2000book} M. A. Nielsen and I. L. Chuang, ``Quantum Computation and Quantum Information'', \emph{Cambridge Uni. Press}, New York, 2000.
\bibitem{Fujiwara:Phy2013} Y. Fujiwara, ``Block Synchronization for Quantum Information'', \emph{Phys. Rev. A}, vol. 87, 022344, 2013.
\bibitem{Bose:1967} R. C. Bose and J. G. Caldwell, ``Synchronizable error-correcting codes,'' \emph{Inf.\ Contr.}, vol.\ 10, pp.\ 616--630, 1967.
\bibitem{Fujiwara:phy2013b} Y. Fujiwara, V. D. Tonchev, and T. W. H. Wong, ``Algebraic techniques in designing quantum synchronizable codes'', \emph{Phy. Rev. A}, vol. 88, 012318, 2013.
\bibitem{Fujiwara:2014} Y. Fujiwara and P. Vandendriessche, ``Quantum Synchronizable Codes From Finite Geometries'', arXiv:1311.3416.
\bibitem{Grassl:2000}  M. Grassl and T. Beth, ``Cyclic quantum error-correcting codes and quantum shift registers,'' \emph{Proc.\ R.\ Soc.\ London Ser.\ A} vol.\ 456, 2689--2706 2000.
\bibitem{MacWilliams:book1978} F. J. MacWilliams and N. J. A. Sloane, ``The theory of error-correcting codes'', North-holland Publishing Comp. 2nd edition, 1978.
\bibitem{Calderbank1996} A. R. Calderbank and P. W. Shor, ``Good quantum error-correcting codes exist,'' \emph{Phys. Rev. A}, vol. 54, pp. 1098-1105, 1996.
\bibitem{Huffman:book2003} W. C. Huffman and V. Pless, ``Fundamentals of Error-Correcting Codes'', \emph{Cambridge Uni. Press}, Cambridge, 2003.
\end{thebibliography}
\end{document}